\documentclass[12pt]{article}
\input epsf.sty

\def\lromn#1{\uppercase\expandafter{\romannumeral#1}}

\usepackage{graphicx,amsmath,amssymb}
\begin{document}
\begin{center}
\begin{large}
\renewcommand{\thefootnote}{\fnsymbol{footnote}}
\textbf{
Toward CP-even Neutrino Beam
}\footnote[4]
{Work supported in part by the Grant-in-Aid for Science Research from
the Ministry of 
Education, Science and Culture of 
Japan No.18654046, No.18684008,  and No. 13135207}

\end{large}
\end{center}

\vspace{1cm}
\begin{center}
\begin{large}
A. Fukumi$^{\dagger}$, I. Nakano$^{\dagger}$,
H. Nanjo$^{\ddagger}$,  N. Sasao$^{\ddagger}$, 
\\ S. Sato$^{\dagger}$, and 
M. Yoshimura$^{\dagger}$

$^{\dagger}$Center of Quantum Universe and
Department of Physics, Okayama University, \\
Tsushima-naka 3-1-1, Okayama
700-8530, Japan
\\
$^{\ddagger}$Department of Physics, Kyoto University,\\
Kitashirakawa, Sakyo, Kyoto,
606-8502 Japan 
\end{large}
\end{center}

\begin{abstract}
%\abst{The best method of measuring CP violating effect in neutrino
The best method of measuring CP violating effect in neutrino
oscillation experiments is to construct and use a neutrino beam made of an ideal mixture of $\bar{\nu}_e$ and $\nu_e$ of monochromatic lines.   The conceptual design of such a beam is described, together with how to measure the CP-odd quantity.   We propose to exploit an accelerated unstable hydrogen-like heavy ion in a storage ring, whose decay has both electron capture and bound beta decay with a comparable fraction.
\end{abstract}

%\kword{neutrino beam, CP violation, neutrino
%oscillation, electron capture, bound beta decay}

%\begin{document}
%\maketitle

%\section{Introduction} 
\vspace{1cm}
It is of great current interest, both to
microphysics and to cosmology, to clarify how
the matter-antimatter asymmetry is generated.
One key idea is the leptogenesis \cite{leptogenesis} in which
the lepton asymmetry of the universe is first generated by
CP violating lepton number violating processes,
and is later converted to the baryon asymmetry
via high temperature B- and L-violating electroweak
processes.
To advance this idea further, it is important
to deepen our understanding on the origin of CP violation in 
the leptonic sector.

Measurement of CP violation in the neutrino
sector is thus one of the most fundamental problems
facing physics beyond the standard particle physics.
In this note we develop a concept of CP-even neutrino
beam, which serves this purpose.
Moreover, it is one of the most
important experimental issues in neutrino physics
to determine the relevant CP parameter in
neutrino oscillation experiments.
\\
%\section{CP-even neutrino beam}
%\label{sec:1}

%\subsection{Concept of CP-even beam}
%\label{sec:2}
The ideal neutrino beam for CP measurement would be a mixture
of monochromatic $\nu_e$ and $\bar{\nu}_e$ beam for which
the detector response is symmetrical, producing after the oscillation equal
numbers of $\mu^{\pm}$ if CP is conserved.
Such a beam is referred to as a CP-even neutrino beam in this paper.
Our proposal is to simultaneously
use the monochromatic neutrino $\nu_e$
of electron capture (EC)
and anti-neutrino $\bar{\nu}_e$ of bound beta ($b\beta$) decay
from hydrogen-like heavy ions.
Such ions do exist, for example \cite{nucleardata}
\begin{eqnarray}
&&
 ^{114}_{49}{\rm In}^{48+}\,,\; 
 ^{110}_{47}{\rm Ag}^{46+}\,,\;
 ^{108}_{47}{\rm Ag}^{46+}\,,\;    
 ^{104}_{45}{\rm Rh}^{44+}\,.
\end{eqnarray}
The concept of CP-even neutrino beam is realized, to a large extent, by this kind of beam.
It is an extension of the ideas of
monochromatic neutrino beam using EC
proposed in \cite{jsato}, \cite{bernabeu}, 
and beta beam proposed in \cite{betabeam}, \cite{betabeamstatus}.

The hydrogen-like heavy ion of this kind has $b\beta$ channels
in addition to continuum beta ($c\beta$) decay and EC, and thus produces
monochromatic $\bar{\nu}_e$ and $\nu_e$ beams once accelerated in a
storage ring.
One can arrange all lines and a part of the continuum neutrino
energy to fall into an optimal range for detection:
above $110 {\rm MeV}$ of the muon production,  but below 
multi-pion production in detector placed at a distance.
We shall argue that this provides an excellent opportunity
of precision experiments to determine the 
CP violation phase $\delta$ \cite{cpmeasurementmethod}
as well as the mixing angle $\theta_{13}$.

Although not exhaustive by any means, a systematic method of search for
candidate ions is briefly described, starting from data of neutral atoms. Since we momentarily ignore $c\beta$ contribution for detection, one important measure
for the CP-even beam is a large EC rate of neutral atom.
This rate is almost the same for the case of the heavy ions.
It is then important to look for calculated $b\beta$ rate 
of about twice of EC rate, since cross sections for
$\nu$ and $\bar{\nu}$ has this ratio (for more details, see
below).
\\

%\subsection{Multi-lines from bound beta decay}
%\label{sec:3}
A possibility of the bound beta decay 
that produces monochromatic
neutrino has been considered theoretically \cite{bahcall},
and their dramatic example that becomes possible only for highly
ionized atoms has been demonstrated 
experimentally \cite{jungetal}.
Relative strength of different neutrino lines is proportional to
the atomic wave function squared at the nucleus.
This factor for the s-wave state of the principal quantum number 
$n$ is $|\psi_{ns}(0)|^2 = (Z/na_B)^3/\pi$ (using for simplicity
the solution of the Schr$\ddot{\mathrm{o}}$dinger equation)
for hydrogen-like atoms of charge $Ze$.
The ratio of the bound to the continuum contribution is given by
\begin{eqnarray}
&&
\hspace*{-0.7cm}
r_B = 
\pi \sum_n N_n (\frac{3.7 keV Z}{n\,Q_{c\beta} })^3
(\frac{Q_{c\beta} + \Delta_{\rm BE}}{Q_{c\beta}})^2\,
K^{-1}(\frac{m_e}{Q_{c\beta}})
\,,
\end{eqnarray}
with $N_n$ the multiplicity factor of available levels and
$\Delta_{\rm BE}$ the difference of binding energy in neighboring ions.
Here  
$K(x)$ 
is related to the phase space integral of the continuum
contribution.
When the effect of Coulomb distortion of wave functions
(the Fermi integral) is ignored,
this function is in the range, 
$K(x) = 0.15 \sim 0.74$ for $x= 1/2 \sim 2$.
When variation of the level difference
$Q_{c\beta} + \Delta_{\rm BE}$ with $n$ is small, the fraction of
$1s$ contribution is $1/(2\zeta(3) -1) \sim 0.71$, $2s$ contribution
$0.18$, and the rest $0.11$.
To obtain a large $b\beta$ rate, $r_B = O[1]$ is required;
for instance, with $Q_{c\beta} + \Delta_{\rm BE} = m_e$, 
the bound ratio $ r_B = 1$ corresponds to $Z = 57$. 
We thus look for heavy atoms of large $Z$.
\\

%\subsection{Beta and EC neutrino spectrum}
%\label{sec:4}
We found the best candidate ion to be $^{108}_{47}{\rm Ag}^{46+}$, and the next 
best one is $^{114}_{49}{\rm In}^{48+}$.
The neutrino and anti-neutrino energy spectrum from decay 
of $^{108}_{47}{\rm Ag}^{46+}$ at rest is shown in Figure 1.
Effect of distorted plane wave under the nuclear Coulomb potential
is important and included in the form of the Fermi integral.

%%%%%%%%%%%%%%%%%%%%%% figure 1 %%%%%%%%%%%%%%%%%%%%%%%%%%%%%%%%%%%%%
%
% For one-column wide figures use

\begin{figure*}[htbp]
 \begin{center}
 \epsfxsize=0.5\textwidth
 \centerline{\epsfbox{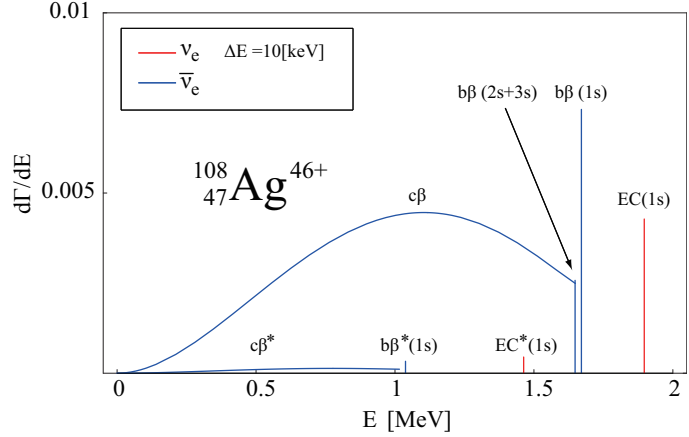}} \hspace*{\fill}
   \caption{Beta and EC neutrino spectrum from $^{108}{\rm Ag}^{46+}$}
   \label{fig:1}  
 \end{center} 
\end{figure*}

%
%%%%%%%%%%%%%%%%%%%%%% figure 1 %%%%%%%%%%%%%%%%%%%%%%%%%%%%%%%%%%%%%
Table 1 gives their monochromatic energies and the endpoint
energy of the continuum, along with their fractional contributions.
To compare the line contribution with the differential continuum
spectrum, we divided the line strength by an energy resolution
factor $\Delta E$, which is taken $10 {\rm keV}$ in Figure 1.
We denoted decay contributions to excited daughter nuclei by
$^*$ in Figure 1 and Table 1.
It is important to use the wave function of the Dirac equation
rather than the Schr$\ddot{\mathrm{o}}$dinger equation,
since the wave function at nucleus is sensitive to
the short distance behavior.
Most conspicuous lines from $^{108}{\rm Ag}^{46+}$ have
fractions; 0.020 adding bound K-,  L- and M-beta decays 
all to the ground nuclear level,
and 0.0096 for $EC + EC^*$. 
%%%%%%%%%%%%%%%%%%%%%% table 1 %%%%%%%%%%%%%%%%%%%%%%%%%%%%%%%%%%%%%
\begin{table}
\caption{Branching ratio of $c\beta$, $b\beta$, and EC from 
$^{108}_{\hspace{0.3em}47}$Ag$^{46+}$ decay}
\label{tab:1}

\begin{center}

\begin{small}
\begin{tabular}{ccc}
\hline\hline
decay mode & neutrino energy $[\mathrm{MeV}]$ & branching ratio  
\\ \hline\hline
$c\beta$                  & $1.649$           & $0.9479$ 
\\
$c\beta^{\ast}$           & $1.016$           & $0.0174$ 
\\ \hline
$b\beta (1s)$             & $1.671$           & $0.0149$
\\
$b\beta(2s)$              & $1.648$           & $0.0041$
\\
$b\beta(3s)$              & $1.645$           & $0.0012$
\\ \hline
$\mathrm{EC(1s)}$         & $1.897$           & $0.0087$
\\ \hline
$\mathrm{EC}^{\ast}(1s)$  & $1.463$           & $0.0009$ 
\\ \hline\hline
\end{tabular}
\end{small}
\end{center}
\end{table}

All promising candidates we know of are listed in Table 2, 
along with their characteristic features.

\hspace{0.5cm}
%\hspace{1zw}

%%%%%%%%%%%%%%%%%%%%%%%%%%-Table2-%%%%%%%%%%%%%%%%%%%%%%%%%%%5%%%%%
\begin{table}
\caption{Neutrino energy (E) and branching ratio (BR) of $c\beta$, $b\beta$ and EC }
\label{tab:2}

\begin{center}

\begin{small}

\begin{tabular}{c|cc|cc}
\hline\hline
ion                            
&\multicolumn{2}{c|}{$^{114}_{\hspace{0.3em}49}$In$^{48+}$} 
&\multicolumn{2}{c}{$^{110}_{\hspace{0.3em}47}$Ag$^{46+}$} 
\\ 
\hline\hline  
half-life                     
&\multicolumn{2}{c|}{70.9[sec]}                            
&\multicolumn{2}{c}{24.4[sec]}                
\\ 
\hline
& $E$[MeV] & BR                           
& $E$[MeV] & BR  
\\ 
\hline
$c\beta$                     
& $1.99$ & $0.981$                        
& $2.89$ & $0.947$ 
\\
$b\beta (1s)$                
& $2.01$ & $0.012$                        
& $2.91$ & $0.005$  
\\
$\mathrm{EC(1s)}$            
& $1.43$ & $0.002$                        
& $0.87$ & $0.001$  
\\ 
\hline\hline 

\end{tabular}

\vspace{0.5cm}

\begin{tabular}{c|cc|cc}
\hline\hline
ion  
&\multicolumn{2}{c|}{$^{108}_{\hspace{0.3em}47}$Ag$^{46+}$}
&\multicolumn{2}{c}{$^{104}_{\hspace{0.3em}45}$Rh$^{44+}$} 
\\ 
\hline\hline  
half-life 
&\multicolumn{2}{c|}{2.36[min]} 
&\multicolumn{2}{c}{42[sec]}  
\\ 
\hline
& $E$[MeV] & BR                           
& $E$[MeV] & BR  
\\ 
\hline
$c\beta$  
& $1.65$ & $0.948$ 
& $2.44$ & $0.971$                
\\
$b\beta (1s)$  
& $1.67$ & $0.015$  
& $2.46$ & $0.006$        
\\
$\mathrm{EC(1s)}$ 
& $1.90$ & $0.009$ 
& $1.12$ & $0.002$                      
\\ 
\hline\hline 

\end{tabular}
\end{small}
%\end{scriptsize}
\end{center}
\end{table}

%%%%%%%%%%%%%%%%%%%%%%%%%%-Table2-%%%%%%%%%%%%%%%%%%%%%%%%%%%5%%%%%

\hspace{0.5cm}
%\subsection{Sensitivity to $\theta_{13}$ and $\delta$,
%and the best location}
%\label{sec:5}
CP violating effects are present only for the appearance experiment
\cite{yy}, hence we need to detect the muon neutrino from
the oscillation, $\nu_e \rightarrow \nu_{\mu}$ and
$\bar{\nu}_e \rightarrow \bar{\nu}_{\mu}$, 
thus to set the neutrino energy above muon production. 
The appearance probability for $\nu_e \rightarrow \nu_{\mu}$ 
(and $\bar{\nu}_{e} \rightarrow \bar{\nu}_{\mu}$)
is to a good approximation given by
\begin{eqnarray}
&&
P_{\nu_e \nu_{\mu} (\bar{\nu}_e \bar{\nu}_{\mu}) } = 
s_{23}^2\sin^2 2\theta_{13}\,\sin^2 \frac{\delta m^2_{13} L}{4E}
\nonumber \\ &&
+ c_{23}^2\sin^2 2\theta_{12}\,\sin^2 \frac{\delta m^2_{12} L}{4E}
\nonumber \\ &&
+ J \cos \left(\pm \delta -  \frac{\delta m^2_{13} L}{4E} \right) 
\frac{\delta m^2_{12} L}{4E}
\sin \frac{\delta m^2_{13} L}{4E} \,,
\label{oscillation probability}
\end{eqnarray}
with $J = c_{13}\sin 2\theta_{12}\sin 2\theta_{23}\sin 2\theta_{13}$.
We note that
for the range of 
$(\delta m^2_{12}/(4 m^2_{13}))^2< \theta_{13} < \delta m^2_{12}/(4 m^2_{13})$,
which implies roughly $\theta_{13} = 0.0025 \sim 0.05$, the last CP sensitive
term in eq.(\ref{oscillation probability}) is the largest.
The matter effect that may mimic CP effects
is negligible at this low energy.
To determine $\theta_{13}$ and $\delta$ with precision, 
choice of the detector location $L$ is important.
Taking account of the neutrino flux factor $\propto 1/L^2$,
three terms in the oscillation probability (\ref{oscillation probability}) 
have different $L-$dependence,
$\propto L^{-2}\,, L^{0}\,, L^{-1} $ respectively,
when the phase $\varphi = \delta m^2_{13} L/(4E)$ is near the oscillation peak.
To maximize simultaneously the flux at the detector and sensitivity to 
CP parameter $\delta$, the best location is at the first peak $\varphi=\pi/2$.
With this choice, the last term in (\ref{oscillation probability})  
becomes proportional to $\pm J \sin \delta$.
Thus the symmetric combination of measured quantities
$P_{\nu_e \nu_{\mu}} + P_{ \bar{\nu}_e \bar{\nu}_{\mu} }$ 
is sensitive to $\theta_{13}$,
while the asymmetric combination 
$P_{\nu_e \nu_{\mu}} - P_{ \bar{\nu}_e \bar{\nu}_{\mu} }$
to the CP violation parameter $J \sin \delta$.
This choice fixes the relation between $E$
and $L$ to be   $L/E  \simeq 517 {\rm km}/1000 {\rm MeV}$.
In Figure 2 we plot the oscillation probability 
$P(\nu_e \rightarrow \nu_{\mu})$
as a function of $\delta$ for three values of 
$\sin^2 2\theta_{13} = 0.1\,, 0.05\,, 0.01$.
\\
%
%%%%%%%%%%%%%%%%%%%%%% figure 2 %%%%%%%%%%%%%%%%%%%%%%%%%%%%%%%%%%%%%
%
\begin{figure*}[htbp]
 \begin{center}
 \epsfxsize=0.5\textwidth
 \centerline{\epsfbox{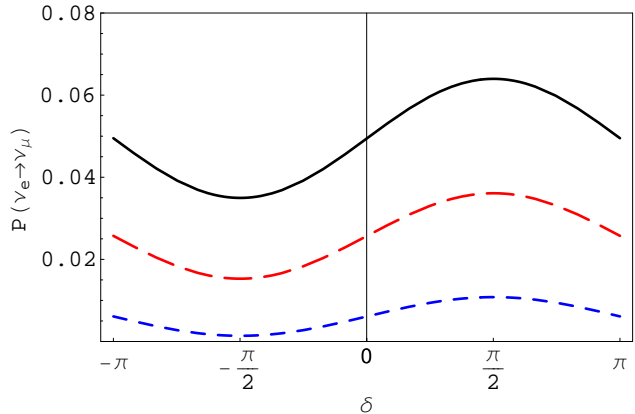}} \hspace*{\fill}
   \caption{Oscillation probability $P_{\nu_e \nu_{\mu}}$ at
  the first peak $\varphi=\pi/2$ for 
  $\sin^2 2\theta_{13} = 0.1\,, 0.05\,, 0.01$ (from top to bottom).
  $P_{ \bar{\nu}_e \bar{\nu}_{\mu} }$ is obtained with 
  $\delta \rightarrow -\delta$.}
  \label{fig:2}  
 \end{center} 
\end{figure*}

%%%%%%%%%%%%%%%%%%%%%% figure 2 %%%%%%%%%%%%%%%%%%%%%%%%%%%%%%%%%%%%%
\hspace{0.5cm}
%\subsection{Optimal choice of beam energy}
%\label{sec:6}
%In our proposed scheme, t
The neutrino energy $E$
can be chosen at will by adjusting the acceleration $\gamma$.
There are many practical factors to be considered; here we select them
mainly from the viewpoint of neutrino detection.
The cross section becomes larger as the energy goes up, but
distinction between $\mu^{+}$ and $\mu^{-}$, mandatory to
establish genuine CP violating effects, 
becomes harder.
Below the single-pion production threshold, backgrounds due to $\pi^{\pm}$'s
are absent; but the cross section is small and the
inevitable Fermi motion smears
out kinematic relation between $\mu^{\pm}$ energy and scattering angle.
Our optimal choice is around $1000$ MeV, which is somewhat above the multi-pion 
production threshold.
At this energy, the cross section for quasi-elastic process
($\nu_{\mu}n \rightarrow \mu^{-}p$ or $\bar{\nu}_{\mu}p \rightarrow \mu^{+}n$)
is $1.5 - 3 $ times as large as for the single-pion production
cross section.
Other channels such as multi-pion productions are negligible.

We note, however, that optimal choice of the neutrino energy
may be changed in actual experiments, depending upon 
chosen technologies 
of accelerators as well as detectors.
\\

%\subsection{Comments on actual beams}
%\label{sec:7}
The concept of CP-even beam is both detector 
and neutrino energy dependent, since
the $\nu_{\mu}$ and $\bar{\nu}_{\mu}$ cross sections are 
different for different targets, and the line
strengths are different at different energies.
One may define the CP-evenness $\eta$,
using a beam flux ${\cal F}(\nu_{e})$ from ions weighted by the cross section
$\sigma(\nu_{\mu})$ at definite neutrino energies;
\begin{eqnarray}
&&
\eta(E\,; \gamma) = \frac{{\cal F}(\nu_e)\sigma(\nu_{\mu}) 
- {\cal F}(\bar{\nu}_e)\sigma(\bar{\nu}_{\mu})}
{{\cal F}(\nu_e)\sigma(\nu_{\mu}) 
+ {\cal F}(\bar{\nu}_e)\sigma(\bar{\nu}_{\mu})}
\,.
\end{eqnarray}
In order to determine $\eta$ for actual beams, we need to know two quantities:
one is the relative intensity of $\nu_{e}$ and $\bar{\nu}_{e}$ flux,
and the other is the relative $\nu_{\mu}$ and $\bar{\nu}_{\mu}$
cross section.
The former quantity, ${\cal F}(\nu_e)/{\cal F}(\bar{\nu}_e)$, 
may be estimated from the measured branching ratio of EC and $\beta$-decay.
As to the latter quantity, $\sigma(\nu_{\mu})/\sigma(\bar{\nu}_{\mu})$, 
we must include  a detector response function (efficiency etc.) in reality.
In this case, the ratio would be best determined in a separate experiment with 
an identical detector component.
As an example, let us 
take $^{108}{\rm Ag}^{46+}$ as a beam, and 
an iron detector.
The uncertainty of ${\cal F}(\nu_e)/{\cal F}(\bar{\nu}_e)$ is $\sim 7\%$ at present \cite{nucleardata} and CP-evenness $\eta \approx 0.09$ selecting the major four lines, neglecting energy dependence of cross sections, and including the effect of Fermi motion. 
If the contribution of $c\beta$ integrated over the energy interval $\approx 8 {\rm keV}$ below the threshold is added, the CP-evenness $\eta \approx 0$.
We have so far discussed the idea of a single-ion beam.
From the point of larger rates, a multi-ion beam is equally interesting.
The multi-ion beam
consists of a simultaneous or 
time sharing circulation of two different ions of
nearly equal $Z/A$'s, each suitable
for the bound beta and EC, individually having higher rates.
For this beam it is important to identify and use ions of
largest rates for $b\beta$ and EC, respectively.
Both candidates of larger $b\beta$ or EC rates than the single-ion
isotopes are numerous.
We only mention two of them;
$^{122}_{48}$Cd$^{48+}$ for $b\beta$, and $^{152}_{70}$Yb for EC.
The event rate of neutrino detection depends
on the isotope factor of $\Gamma /Q$ 
with $\Gamma$ the partial $b\beta$ or EC decay rate.
The isotope $^{122}_{48}$Cd$^{48+}$, for instance, has the 1s $b\beta$
event rate larger by a factor $\approx 12$ than the $b\beta$ rate 
of $^{108}_{47}$Ag$^{46+}$,
and $^{152}_{70}$Yb EC rate is larger by $\approx 600$ than
the corresponding EC rate of $^{108}_{47}{\rm Ag}^{46+}$.
The mixed beam made of $^{122}_{48}$Cd$^{48+}$ (for $\bar{\nu}_e$)
and $^{152}_{70}$Yb$^{60+}$ (for $\nu_e$) has similar $Z/A$,
and it may be possible to accelerate both of them simultaneously.
Their monochromatic neutrino energies are
$3.0 $MeV ($1s$ $b\beta$) for $^{122}_{48}$Cd$^{48+}$
and $5.0 {\rm MeV} $ ($1s$ EC) for $^{152}_{70}$Yb,
hence the time sharing circulation might be more appropriate.
This option of multi-ion beam opens more varieties
toward the CP-even beam and should be kept in mind for further study.
\\

%\textit{Note added}:
After submitting this paper, %for publication
%it was pointed out by an anonymous referee
we have noticed
that the measured EC decay rate  of hydrogen-like $^{140}$Pr$^{58+}$
ion is about 50$\%$ larger than that of helium-like $^{140}$Pr$^{57+}$
ion \cite{litvinov}, 
due to angular momentum restriction of both
leptons and nuclei. \cite{folan} \cite{patyk}
A similar mechanism may occur in our system $^{108}$Ag$^{46+}$,
$^{110}$Ag$^{46+}$,$^{114}$In$^{48+}$
thereby enhancing our EC rates of Tables 1 and 2.
\\

%\section{Summary}
%\label{summary}
In summary,
we proposed a new 
concept to implement the 
experimental method optimal for determination of the CP parameter
of neutrino mixing
by using monochromatic neutrino beams both from electron
capture and bound beta decay. 
Feasibility of such beams hinges crucially on production and 
storage of high intensity unstable nuclear beams, and innovative works 
towards this direction are needed.
\\

%\textit{Note added}:
%After submitting this paper for publication
%it was pointed out by an anonymous referee
%that the measured EC decay rate  of hydrogen-like $^{140}$Pr$^{58+}$
%ion is about 50$\%$ larger than that of helium-like $^{140}$Pr$^{57+}$
%ion \cite{litvinov}, 
%due to angular momentum restriction of both
%leptons and nuclei \cite{folan} \cite{patyk}.
%A similar mechanism may occur in our system $^{108}$Ag$^{46+}$,
%$^{110}$Ag$^{46+}$,$^{114}$In$^{48+}$
%thereby enhancing our EC rates of Tables 1 and 2.
%\\

%\newpage
%\vspace{1cm}
%\textit{Acknowledgements}:
%We should like to thank the anonymous referee for
%kindly bringing our attention to the work of ref. 11) and suggesting its relevance to our work.

\end{document}